\documentclass[preprint]{nature}
 
\usepackage{hyperref}
\usepackage{movie15}
\usepackage{graphicx} 
\usepackage{epsfig} 

\title{Morphology, cell division, and viability of \emph{ Saccharomyces cerevisiae} at high hydrostatic pressure}

\author{Khanh Nguyen\textsuperscript{1}, Steven Murray\textsuperscript{1} Jeffrey A. Lewis\textsuperscript{2}, Pradeep Kumar\textsuperscript{1*} } 
 
\begin{document} 
 
\maketitle 
 
\begin{affiliations} 
\item Department of Physics,  University of Arkansas, Fayetteville, AR 72701 USA

\item Department of Biological Sciences, University of Arkansas, Fayetteville, AR 72701 USA

\end{affiliations}

\begin{abstract}

High hydrostatic pressure is commonly encountered in many environments, but the effects of high pressure on eukaryotic cells have been understudied. To understand the effects of hydrostatic pressure in the model eukaryote, \textit{Saccharomyces cerevisiae}, we have performed quantitative experiments of cell division, cell morphology, and cell death under a wide range of pressures. We developed an automated image analysis method for quantification of the yeast budding index - a measure of cell cycle state - as well as a continuum model of budding to investigate the effect of pressure on cell division and cell morphology. We find that the budding index, the average cell size, and the eccentricity - a measure of how much the cell morphology varies from the being elliptical - of the cells decrease with increasing pressure. Furthermore, high hydrostatic pressure led to the small but finite probability of cell death via both apoptosis and necrosis. Our experiments suggest that decrease of budding index arises from cellular arrest or death at the cell cycle checkpoints during different stages of cell division.
\end{abstract}

\section*{Introduction}
Both prokaryotic and eukaryotic microbes thrive under a staggering array of diverse environmental conditions~\cite{Yayanos1981, Kato1998, Brock1969, Bakermans2003, Schleper1995, Horikoshi1982} including high pressure and extremes of temperature. High pressure and temperature present biophysical challenges to cells, thus, cellular growth and survival of organisms living under these extreme conditions are likely maintained by subtle adaptations of their biomolecular machinery, and hence optimizing molecular function for each particular environmental challenge. Extremes of pressure and temperature are known to affect the stability and functionality of proteins~\cite{Privalov1990, Mozhaev1996, Hawley1971}. Many proteins are known to denature at a pressure above 200MPa due to decreasing hydrophobicity of proteins with pressure~\cite{Gross1994, Hummer1998}. While moderate pressures may not lead to a complete denaturation of proteins; they may affect the functionality of proteins~\cite{Merrin2011, Kumar2013}. Although the effects of pressure on physicochemical basis is well defined, the effects on the living organisms have been elusive due to complexities of the organisms~\cite{Abe2004}.

Multiples studies on mesophilic organisms have suggested that high pressure can cause growth inhibition, cellular arrest,  as well as cell death~\cite{Sonoike1992, Kato1998, Seki1998}. Recent studies suggest that elongation of \textit{Escherichia coli (E. coli)} at high pressure results from depolymerization of a key cytoskeletal proteins FtsZ~\cite{Ishii2004} and MreB~\cite{Kumar2016}. Furthermore, these studies also reveal the heterogeneous stochastic nature of cell division at high pressures~\cite{Kumar2013,  Huh2011}. The elongated cells obtained up to 70MPa revert to their normal length once the pressure is decreased to atmospheric pressure suggesting that the cellular processes retain their normal functionality over a short time once the high pressure is released~\cite{Kumar2016}. Nishiyama et. al. using a high-pressure imaging chamber have shown that increasing pressure results in the decrease of motility of cells with a complete inhibition of motility at 80MPa~\cite{Nishiyama2012}.  Spectroscopy and imaging techniques revealed that very high pressure could damage cell membrane of \textit{E. coli}. Study on application of very high pressure on \emph{Escherichia coli} suggest that the cells can be killed above 550 MPa \cite{Sonoike1992}, while extremophilic prokaryotes isolated from the Mariana Trench can grow well above pressure 100 MPa \cite{Kato1998}. On the contrary, eukaryotic organisms tend to be more sensitive to pressure, with the exception of tardigrades (\emph{Milnesium tardigradum}), which can survive over 600 MPa\cite{Seki1998}.

To understand and model the effects of high pressure in eukaryotes, we chose the budding yeast \textit{Saccharomyces cerevisiae} due to the wealth of genetic tools and physiological knowledge. Yeast is a unicellular eukaryotic organism belonging to Fungi kingdom. A typical yeast cell is around 5-10 micrometers in diameter, and the cells reproduce through a process called budding. Budding is a form of asymmetric asexual cell division; when once a ``mother" cell reaches a critical size, it gives rise to a ``daughter" cell (the bud) made of an entirely new surface. A bud appears early in the cell cycle (G1 phase), and the site at which the bud is formed becomes a channel that connects the mother and daughter cell so that the nucleus and other organelles can pass to the daughter cell \cite{Hartwell1974, Johnston1977, Hartwel1974}. The replicative lifespan of \emph{S. cerevisiae} is as short as 2 days, which makes it an excellent laboratory candidate for studying the effect of high hydrostatic pressure on the growth and the viability of yeast \cite{Shibata2014, Palou1998, Fernandes2005}.

High hydrostatic pressure has been shown to affect the activities of \emph{S. cerevisiae} of the certain strains up to 300 MPa \cite{Gross1994, Lammi2004}. For example, high pressure can induce loss of mitochondrial function \cite{Rosin1977}, chromosome abnormalities including polyploidy \cite{Hamada1992, Bravim2014, Fernandes2004}, and it disrupts the ultrastructure of the cells \cite{Kobori1995}. It also has been found that a short duration of heat shock at $43^oC$ for 30 minutes increases the yeast's tolerance to pressure and allows the cells to survive up to 150 MPa \cite{Iwahashi1991}. Recent studies suggest protein synthesis is entirely shut down at 67 MPa \cite{Gross1994}, and the protein denaturation occurs in the pressure range of 100-300 MPa.

While other studies have explored the effect of high pressure on \emph{S. cerevisiae}, a quantitative understanding of the effects of long exposure to high pressure is still missing. Here we sought a quantitative study of the effects of high hydrostatic pressure on the cell division, cell morphology, and cell viability of \emph{S. cerevisiae}.  We subjected \emph{S. cerevisiae} to a wide range of high hydrostatic pressures and quantified the cell size, cell morphology, cell division, and budding index. Our results suggest that high pressure suppresses the budding of \emph{S.cerevisiae} leading to a decreased budding index consistent with G1 cell cycle arrest.  Furthermore, we also investigated the nature of cell death upon long exposure to high pressures.  We find that yeast cells exposed to high pressures for an extended period can undergo early stages of apoptosis, as well as necrotic cell death. Moreover, we develop a continuum phenomenological model to account for the high-pressure effects on the cell division.

\section*{Materials and Methods}

\noindent{\bf Cell culture and media}

A fully prototrophic haploid lab derivative of DBY8268 (S288C background) \cite{Kim2007} of {\it S.cerevisiae} was used for all the experiments reported here. A small amount of cells were first streaked out onto a petri dish containing agar-YPD ($1.5\%$ agar, $1\%$ yeast extract, $2\%$ peptone, $2\%$ dextrose)~\cite{Saghbini2001} and were allowed to grow for 48 hours. A single colony was picked from the petri dish and transferred into a tube containing 10 ml of liquid YPD media. The cells were then grown in an incubator at $30^{\circ}$C for $24$~hours.  After the incubation,  50 $\mu$L of the liquid culture was added to 1200 $\mu$L of YPD media and was subsequently transferred into the high-pressure cuvette for pressure experiments.

\noindent{\bf Propidium iodide and Annexin staining}

To investigate the cause of cell death,  yeast cells were stained using Propidium Iodide (PI) and Annexin V (AnnV). PI and AnnV conjugated with Alexa Fluor 488 were purchased from Thermofisher, USA. After subjecting the cells to high pressures, cells were washed with Phosphate Buffered Saline (PBS) 1X $5$ times and were resuspended in 1X Annexin-binding buffer (Thermofisher, USA).  AnnV and PI were then added to the solution and the sample was incubated at room temperature for 15-20 minutes. After the staining,  cells were washed with 1X Annexin binding buffer 5 times and the phase-contrast, as well as fluorescent images of  the cells were recorded using a Nikon Optiphot 2 microscope.

\begin{figure}[htb]
	\begin{center}
		\includegraphics[width=13cm]{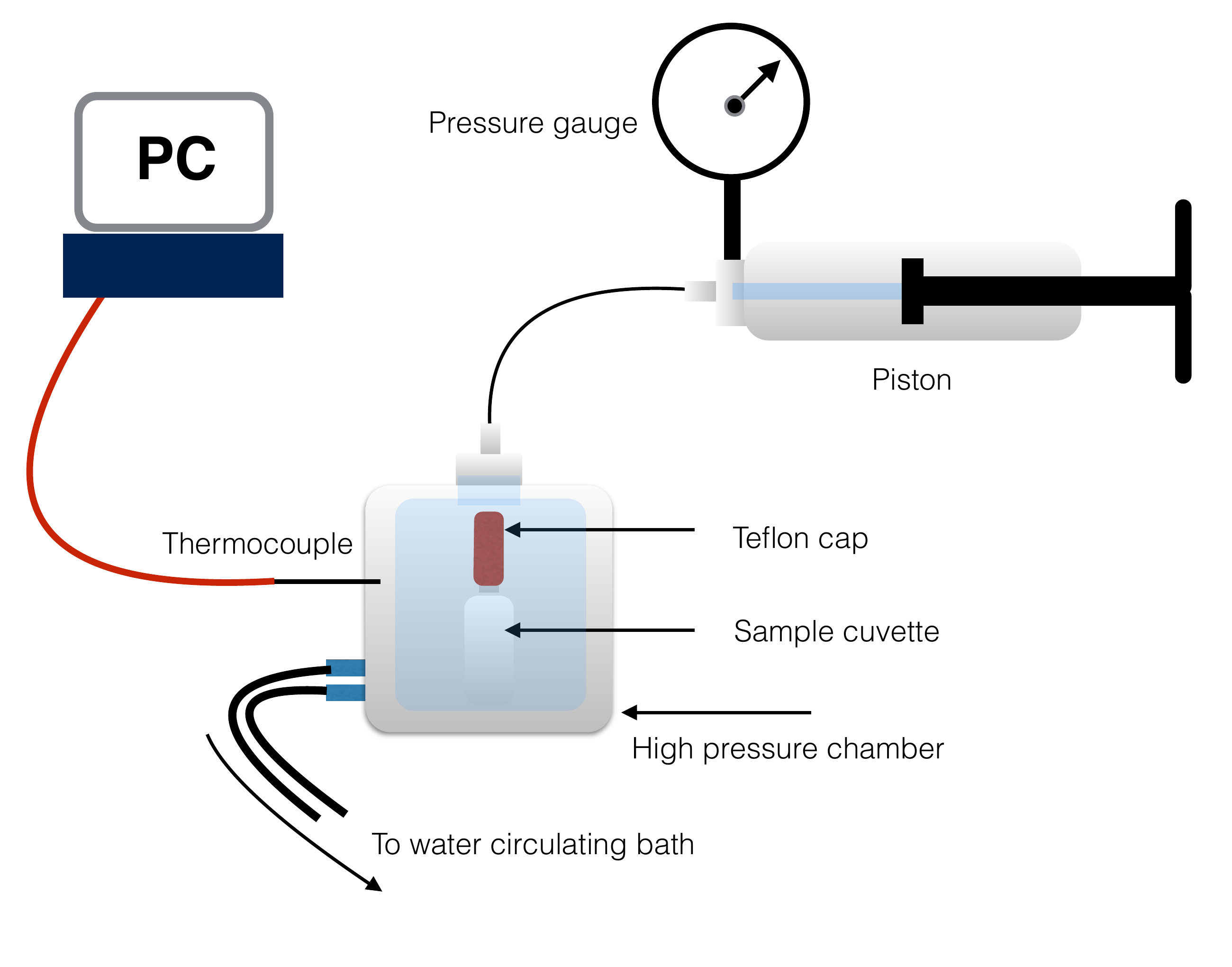}
		\caption{Schematic of the temperature regulated high-pressure experimental setup.}
		\label{fig:setup}
	\end{center}
\end{figure}
\noindent{\bf Experimental setup}

The experimental setup for temperature regulated high-pressure experiments is shown in Fig.\ref{fig:setup} \cite{Kumar2013}. A sample in a cylindrical cuvette (Spectrocell; volume: 1200 $\mu$L) with a removable Teflon cap is loaded into the high-pressure cell (ISS, Illinois USA). A piston (HIP Inc., Pennsylvania, USA) is used to pressurize the water inside the pressure cell, and the pressure is measured using a pressure gauge attached to the piston. The temperature of the sample is maintained using a circulating water bath (NesLab, USA) and it is measured using a thermocouple (National Instruments, USA) connected to the high pressure chamber and is recorded using a data acquisition card (National Instruments, USA) mounted on a PC. The equilibrium temperature fluctuations are $\pm 0.2^o$C, and the pressure uncertainty is estimated to be 1 MPa.  All the experiments at high pressures were performed at temperature $T=30^{\circ}$C. Since temperature equilibration is slower than the pressure equilibration; we first equilibrated the temperature of the pressure chamber to the desired temperature after which the cuvette with the sample is loaded into the pressure chamber. After loading the sample into the pressure chamber, the piston is used to achieve the desired pressure for the experiment.

\noindent{\bf Quantification of cell-size distribution, cell morphology, and budding index}

\begin{figure}[htb]
	\begin{center}
		\includegraphics[width=13cm]{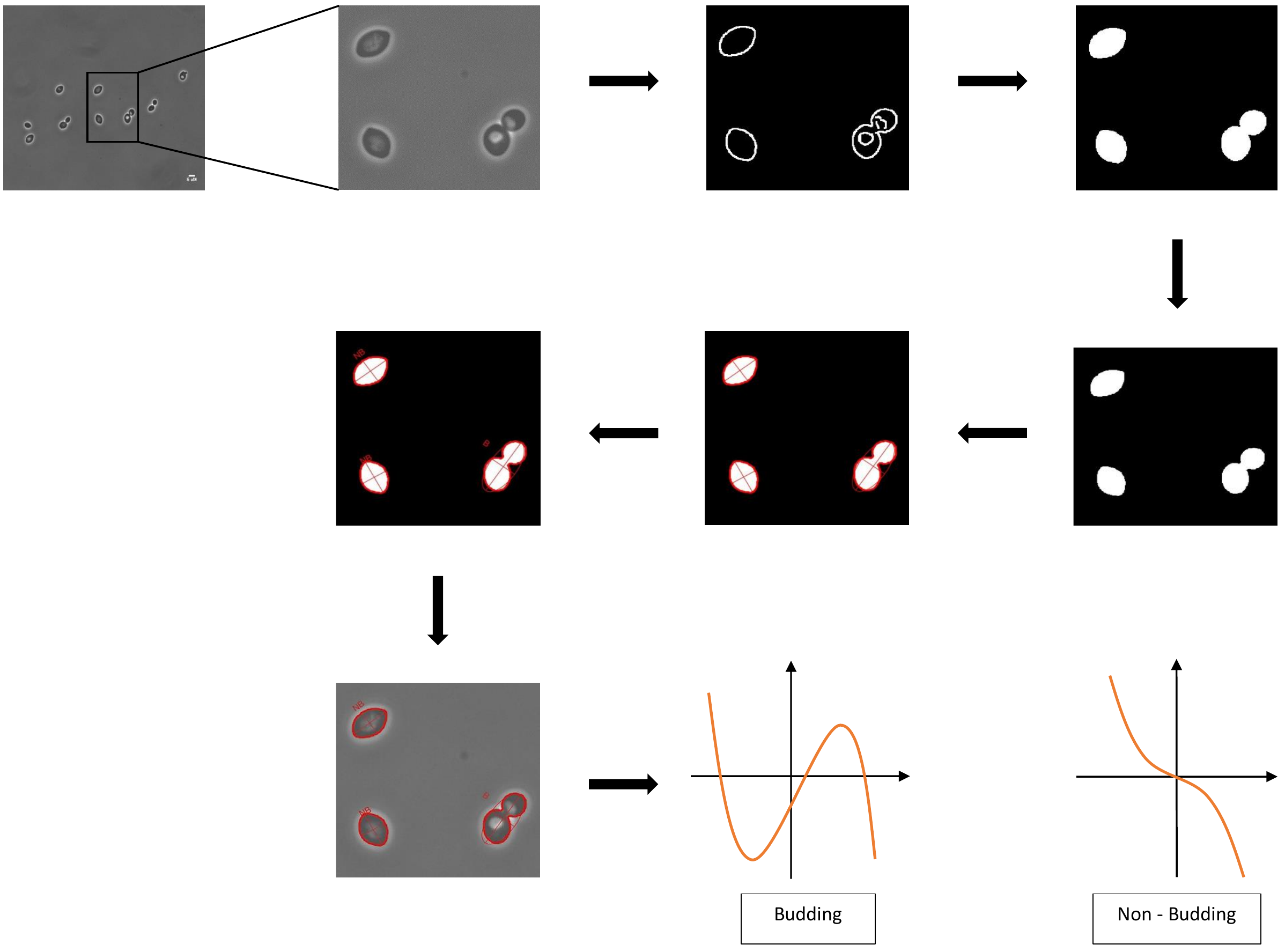}
		\caption{Quantification of budding and non-budding cells. Initial phase contrast images are converted to binary images and are segmented for cell identification. After the identification of the cells, each cell in an image is fitted with an ellipse. The cells are rotated such that the semi-major axis of the ellipses align with the $x$-axis. The local curvature of the periphery of the cells is plotted against $x$. The budding and non-budding cells are identified by a number of zero-crossings of the local curvature of the boundary of the cells.}
		\label{fig:method}
	\end{center}
\end{figure}

In order the quantify the budding index of the sample at different pressures, phase contrast images of the cells were acquired using a $40$X objective and Spot Imaging camera mounted on a Nikon Optiphot 2 microscope. The acquired images were analyzed using a Matlab code for the determination of budding index as well measures of cell morphology such as circumference and eccentricity of the cells. A detailed flowchart of our method of image analysis is shown in Fig.~\ref{fig:method}. Multiple phase contrast images of the cells were first acquired. Images were then processed and converted to binary images and segmented for the identification of cells. Once the cells were identified, each cell was then fitted with an ellipse using an open source Matlab code ~\cite{Gal2003}. The coordinates of each cell in an image were then rotated such that the semi-major axis of the ellipse lied on the $x$-axis. The local curvature of the periphery of the cells was obtained by a moving $n$-point linear fit to the boundary points  as a function of $x$, where the value of $n$ (typically about 5) was varied to optimize the precision of the obtained local curvature. For each cell, the local curvature along the periphery of the cell was then plotted as a function of $x$ and was analyzed for the determination of whether the cell is budding or not. A non-budding cell will only have one change in the sign of its derivative, positive to negative, over the semi-major axis while a budding cell will have more than one.

\section*{Results}

\noindent {\bf Cell-size distribution and budding index of \emph{Saccharomyces cerevisiae} at atmospheric pressure}

\begin{figure}[htb]
	\begin{center}
		\includegraphics[width=12cm]{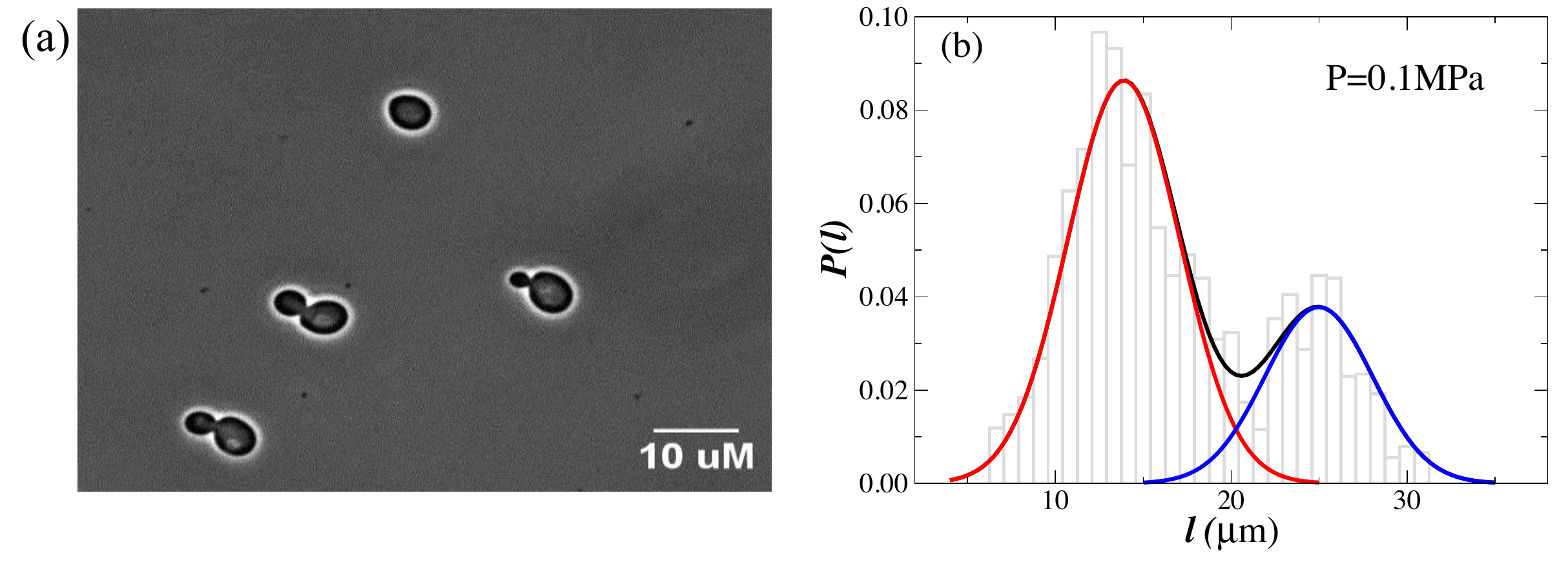}
		\caption{(a) Phase contrast image of  \emph{S.cerevisiae} at a pressure of $0.1$MPa and a temperature of $30^oC$. (b) Probability distribution of the cell size as measured by the circumference of the cells. The bars are the experimental data, and the solid black curve is a fit to the data using a sum of two Gaussian functions shown as solid red and green curves respectively. The areas under the two Gaussian curves are the fractions of the non-budding and budding cells respectively.}
		\label{fig:figCNP}
	\end{center}
\end{figure}

To investigate the effect of pressure on the cell division of \emph{S. cerevisiae},  we first performed a control experiment  to determine the cell size distribution and the budding index of the cells at pressure $P=0.1$ MPa. In Fig. \ref{fig:figCNP}, we show the probability distribution, $P(\ell)$, of the cell size as measured by the circumference of cells, $\ell$, at atmospheric pressure. We find that the distribution of cell size is bimodal and can be fit very well with a sum of two Gaussians centered at the mean values of circumference corresponding to the budding and non-budding cells. The area under the two Gaussian curves are the fractions, $f_{NB}$ and $f_{B}$, of the non-budding and budding cells respectively. 
The ratio $f_{B}/f_{NB}$ is known as the budding index. We find that the budding index of the strain used in our experiments is $\approx 0.463$ at a pressure of $0.1$~MPa and a temperature of $30^{\circ}$~C. Since high pressure can affect the budding index by affecting the timescales of different  cellular processes as well as the cell morphology, we next investigate the effect of high pressures on the budding index and the cell-size distribution.

\noindent{\bf Effects of hydrostatic pressure on the cell-size distribution and budding index of \emph{S.cerevisiae} }

\begin{figure}
	\begin{center}
		\includegraphics[width=12cm]{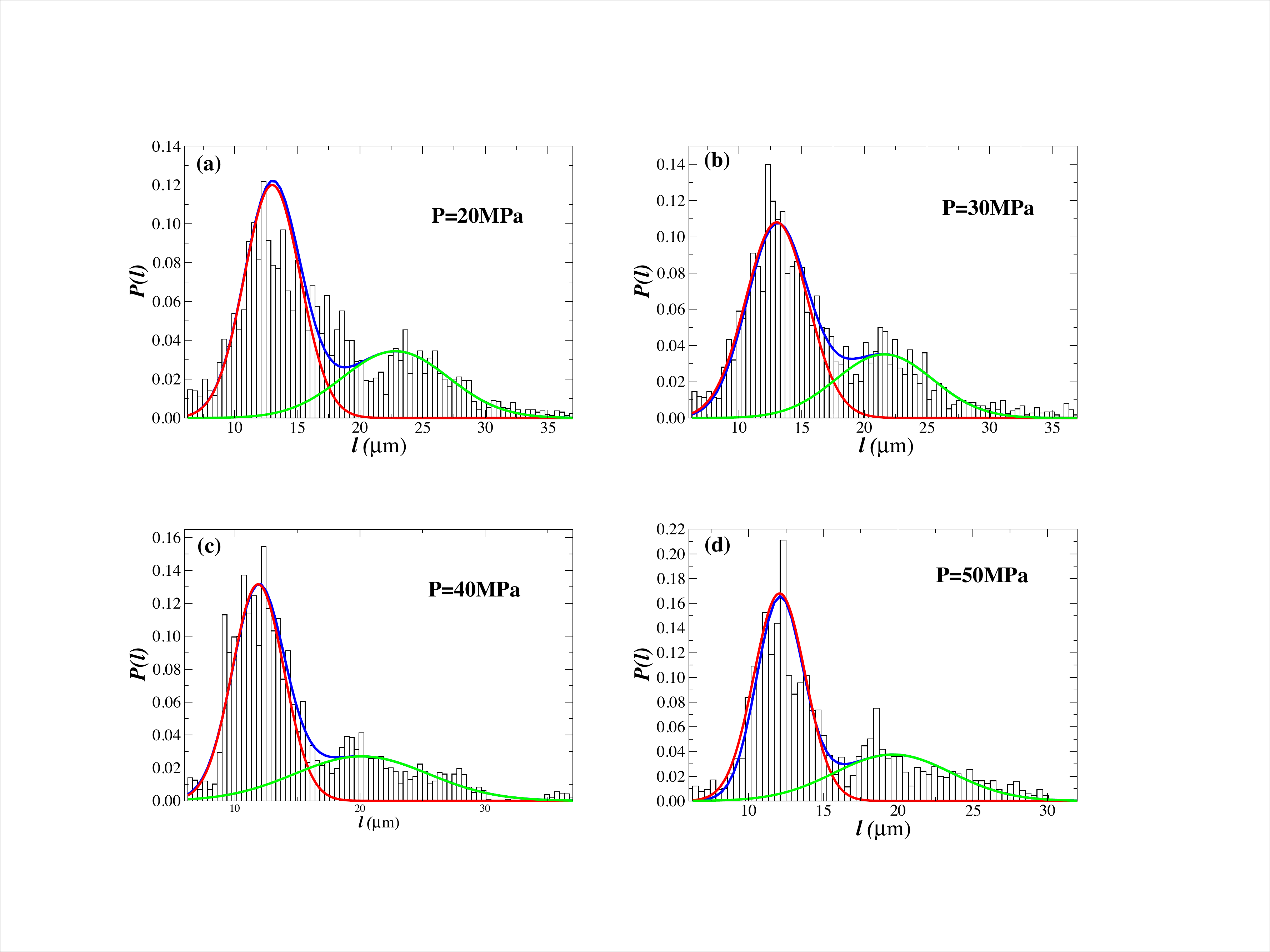}
		\caption{Probability distribution of cell size, as measured by the circumference of the cells, at  $T=30^{\circ}$C and pressures  (a) $20$~MPa. (b) $30~$MPa, (c) $40$~MPa, and (d) $50$~MPa. The distribution of cell size can be fit well with a sum of two Gaussian functions (shown as the solid black curve). The individual Gaussian functions are plotted as solid red and green curves respectively. The area under the red Gaussian curve is the fraction of non-budding cells while the area under the green solid curve is the fraction of budding cells.}
	\label{fig:HOR2press}
	\end{center}
\end{figure}

To determine the effects of high pressure on \emph{S. cerevisiae}, we performed high pressure experiments for four different pressure: $P=20$ MPa, $P=30$ MPa, $P=40$ MPa, and $P=50$ MPa. The experiments at high pressures were carried out until the cell growth reaches the saturation. Since the growth rate of cells depend on the pressure, the total time over which the pressure was applied to the cells varied with pressure. For pressure $P=0.1$MPa, the duration was $12$~hours, while for $P=20$ MPa and $30$~MPa, we ran the experiments for $24$~hours and for the higher pressures, $P=40$ MPa and $P=50$MPa, the durations of experiments were $48$~hours. After exposing the cells to these pressures, the samples were taken out of the high pressure chamber and imaged immediately for the determination of cell morphology and budding index. For the statistical significance of the experimental data, multiple images of the cells were acquired and the statistics were gathered for more than $4000$~cells for each pressure.

The probability distribution of cell size of \emph{S.cerevisiae}, $P(\ell$), as measured by the circumference of the cells $\ell$,  as a function of $\ell$, is shown in Fig.\ref{fig:HOR2press} for four pressures (a) $20$~MPa, (b) $30$~MPa, (c) $40$~MPa, and (d) $50$~MPa. Similar to the probability distribution of $l$ for $P=0.1$~MPa (see Fig.~\ref{fig:figCNP}), the distribution of all the pressures can be reasonably fit with a sum of two Gaussian functions (shown as solid black curves). The individual Gaussian functions are also shown as solid red and green curves for all the pressures. The area under the red curves and green curves are the fractions of non-budding and budding cells respectively. 

\begin{figure}
	\begin{center}
		\includegraphics[width=10cm]{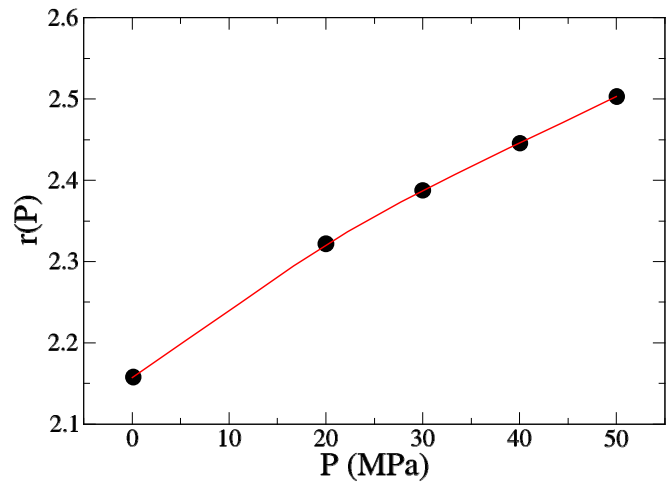}
		\caption{The ratio $r$ of non-budding to budding cells as a function of pressure. Red circles are the data, and the solid black curve is a linear fit through the experimental data points. The value of $r$ increase as pressure increases, suggesting that the budding index decreases upon increasing pressure.}
	\label{fig:r_press}
	\end{center}
\end{figure}

Figure~\ref{fig:r_press}, we show the ratio of the fractions of non-budding to budding cells, $r=f_{NB}/f_B$ as a function of pressure. We find that $r$ increases with pressure suggesting that the budding index decreases with increasing pressure. In other words, the budding index is a function of pressure and decreases upon increasing pressure.

\noindent{\bf Effects of hydrostatic pressure on the cell morphology of \emph{S.cerevisiae} }

\begin{figure}
	\begin{center}
		\includegraphics[width=14cm,height=14cm]{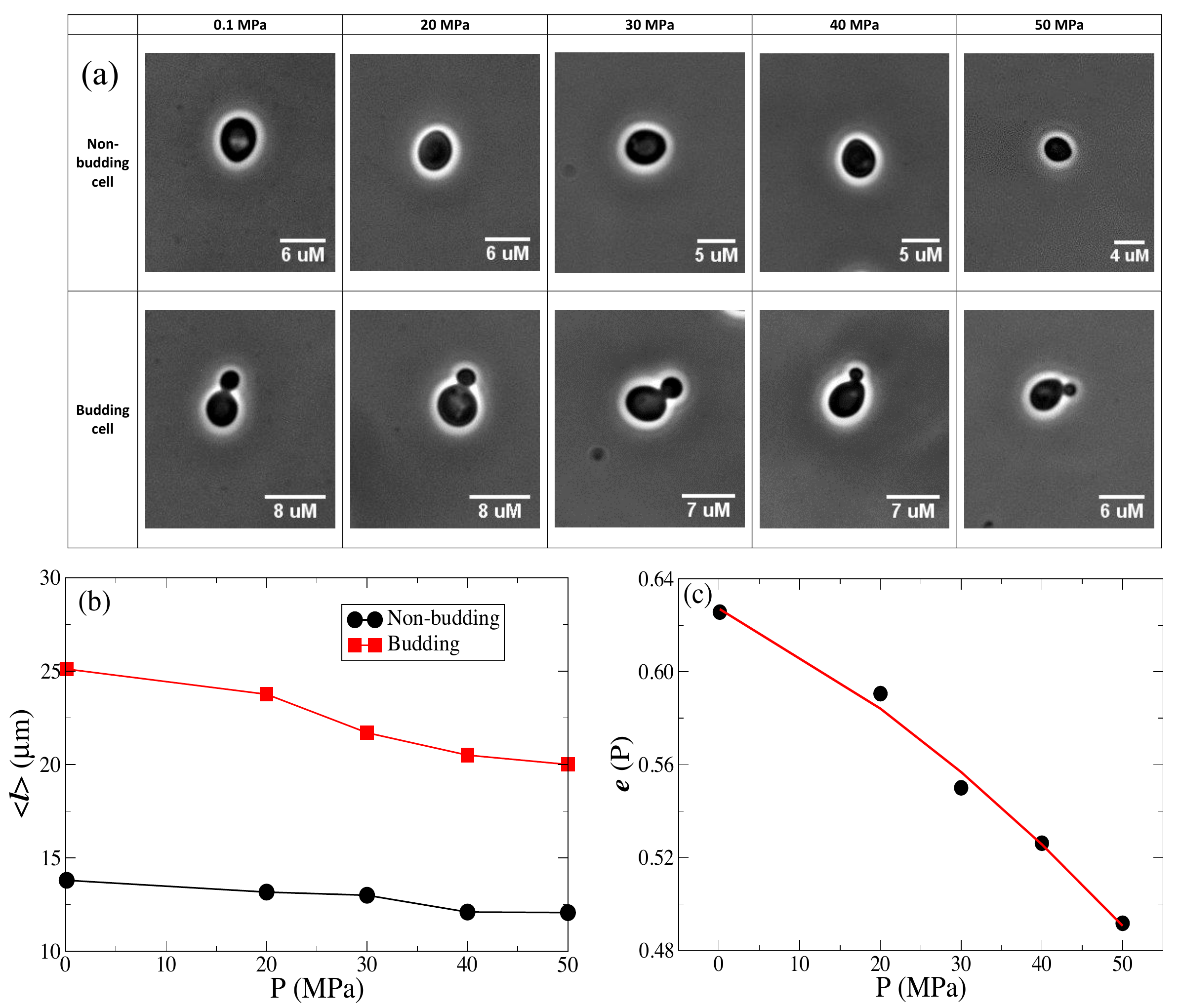}
		\caption{(a)Representative morphology of cells after exposure to different pressures. The cells tend to become smaller and more circular with increasing pressure. (b) Average cell size $\left < \ell \right>$ as a function of pressure. $ \left< \ell \right>$ decreases with increasing pressure.  (c) Average eccentricity, \textit{e}(P) of the non-budding cells. Eccentricity decreases upon increasing pressure, suggesting that the cells become more circular. }
		\label{fig:morph}
	\end{center}
	
\end{figure}

We next investigated the effect of high pressure on the morphological changes in \emph{S.cerevisiae}. The essential morphology in the classification and identification of yeasts includes the description of shape, size, and internal structure of yeast cells; the differences during reproduction of cells and the position of the daughter cells to the parents; the changes when the cells undergo sexual activities. The shape of yeast cell and its structural parts under the microscope can be viewed as two-dimensional objects, such as circular, elliptical or bottle shaped \cite{Suzuki2004, Becze1995, Knop2011}. The standard \emph{S.cerevisiae} are elliptical, or occasionally spherically shaped with a typical diameter of 4 to 8 $\mu$m for spheres, and about 7 $\mu$m along the semimajor axis for elliptical cells \cite{Becze1995}. To quantify the morphology of cells exposed to different pressures, we measured average cell size and eccentricity of the cells after exposing the cells to different pressures. In Fig. \ref{fig:morph} (a), we show a representative image of the morphological changes in the cells as a function of pressure. We find that the cells become progressively smaller and circular in shape as the pressure is increased.

In Fig. \ref{fig:morph} (b), we show the average size, $\left<\ell\right>$, of the budding and non-budding cells as determined from the distribution of the cell sizes (Fig. \ref{fig:HOR2press}). We find that average size of both the budding and non-budding cells decrease slightly upon increasing pressure. Moreover, we find that cell-shape changes from more elliptical to circular upon increasing pressure (shown in Fig. \ref{fig:morph} (c)). The cell shape change of the cells at high pressure may arise due to changes in osmolarity regulation, a topic that we will explore in the future.

\noindent{\bf Cell cycle arrest and cell viability at high pressures}

\begin{figure}
	\begin{center}
		\includegraphics[width=12cm]{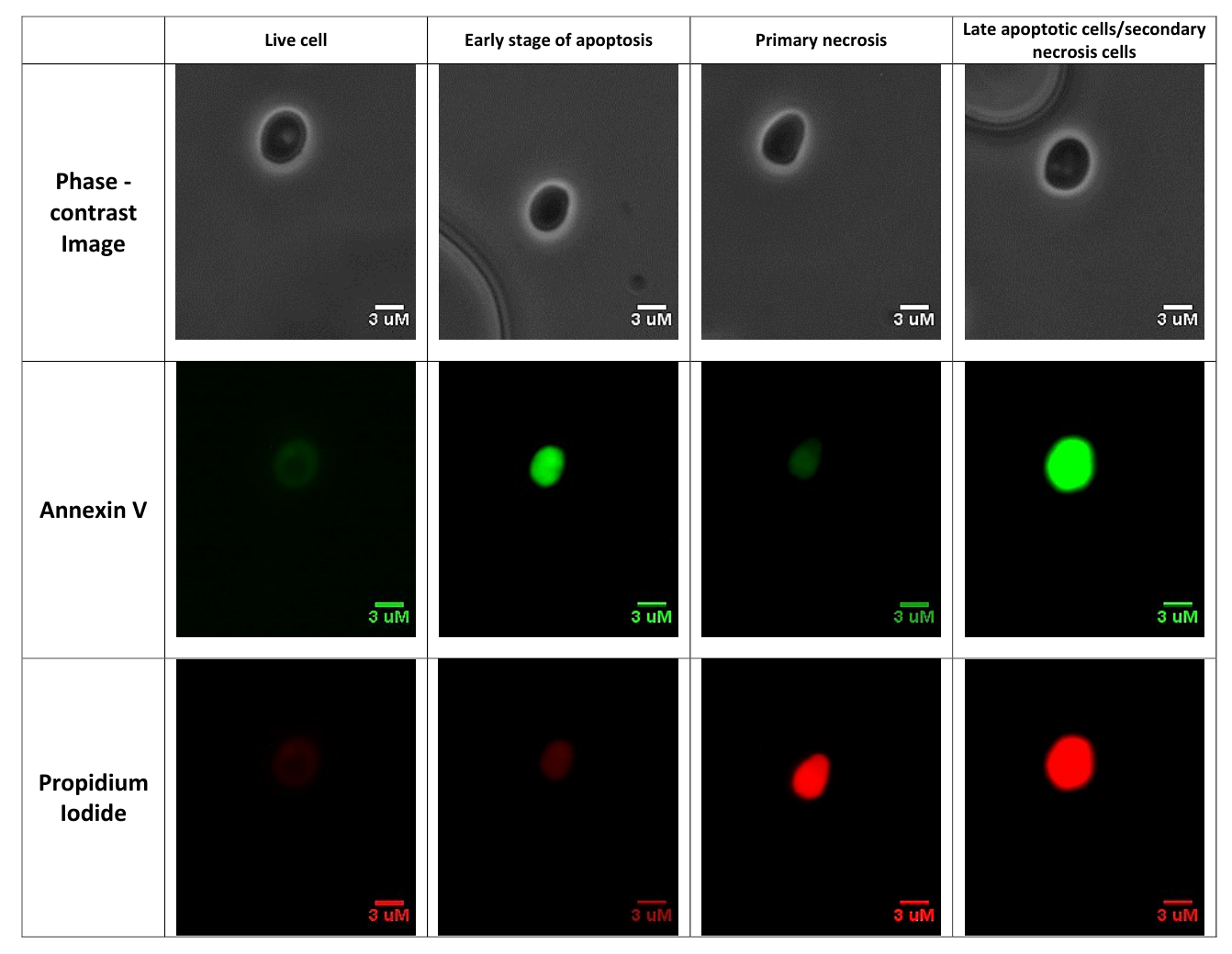}
		\caption{Co-staining of annexin-V (AnnV) and propidium iodide (PI) allows discrimination between live cells ($AnnV^-, PI^-$), early apoptosis cells showing phosphatidylserine (PS) ($AnnV^+, PI^-$), primary necrosis due to membrane rupture ($AnnV^-, PI^+$), and late apoptotic/secondary necrosis cells ($AnnV^+, PI^+$).}
		\label{fig:AnnPi}
	\end{center}
\end{figure}
Our results on the effect of high pressure on budding index, as discussed above, suggest that the likelihood that non-budding cells undergo a complete budding cycle decreases with increasing pressure. This could be caused by either cell cycle arrest or cell death. To check if cell death was the primary reason for the decreased budding index at high pressure, stained cells with annexin-V (AnnV) and propidium iodide (PI). AnnV staining detects phosphatidylserine 	in membranes, which is a conserved apoptotic marker \cite{Madeo1997}. PI is a nuclear dye that only stains membrane compromised necrotic cells, but not apoptotic cells. The identification of each stage is shown in Fig.\ref{fig:AnnPi}. We could identify 4 different types of cells after the staining process; live cells when both AnnV and PI are negative (weak AnnV staining of the cellular membrane and weak PI staining of the nucleus), early stage apoptotic cells when AnnV is positive, and PI is negative, primary necrotic cells due to ruptured membranes when AnnV is negative, and PI is positive, and lastly, both AnnV and PI positive dead cells, which we interpret as late apoptotic \cite{Gutierrez2010, Eisenberg2010, Madeo1997, Buttner2006, Gutierrez2009}. Our results suggest low cellular death rates under the pressure tested, though cell death does increase with increasing pressure. Most of the observed cell death was necrotic, consistent with the previous finding that pressure can cause membrane rupture \cite{Zong2006}.

\noindent{A continuum model of \emph{S.cerevisiae}'s cell division to account for the budding index at normal and high pressures}

\begin{figure}
	\begin{center}
		\includegraphics[width=13cm]{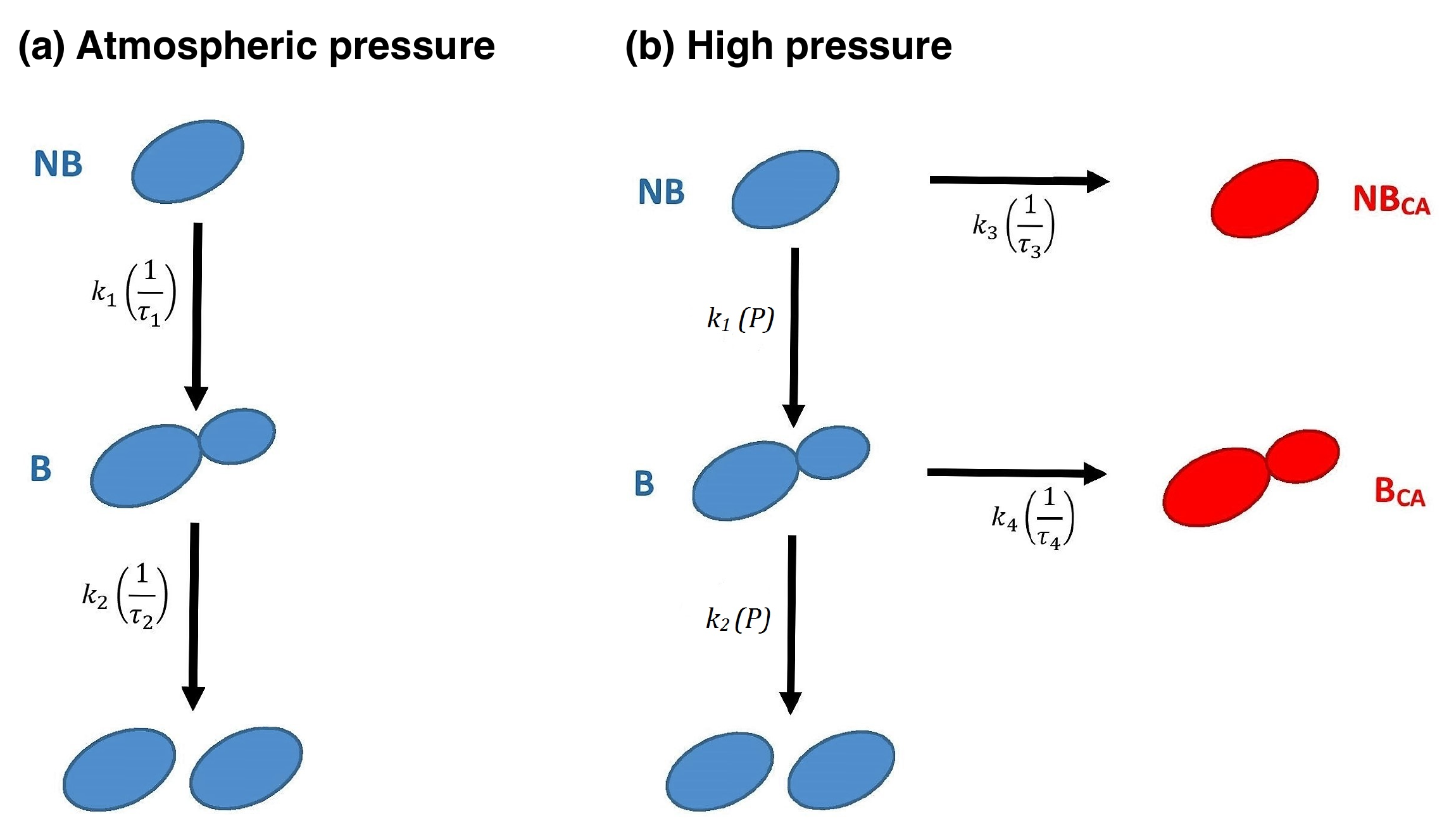}
		\caption{Schematic of the model of cell division at normal and high pressure.}
		\label{fig:model}
	\end{center}
\end{figure}

At normal pressure, cell division of budding yeast begins with the growth of non-budding ``mother" cells. When a mother cell reaches a certain size early in G1 phase, a ``daughter" bud emerges. The mother and daughter cells remain connected by a cytoplasmic channel that allows the nucleus and other organelles to pass through from mother cell to daughter cell \cite{Hartwell1974}. A schematic of a cell cycle process is shown in Fig.\ref{fig:model}. We assume that the rate constant of the initial growth of the non-budding cell is $k_1$ which relates to the time-scale of initial growth $\tau_1$. When the bud is formed, the mother and daughter cells are formed over the time-scale $\tau_2$ with the corresponding rate constant $k_2$. Following the argument above, the growth kinetics of budding yeast at normal pressure can be written as
\begin{equation}
\frac{dNB}{dt} = -k_1 NB + 2k_2 B
\label{eqn:eqn1}
\end{equation}
\begin{equation}
\frac{dB}{dt} = k_1 NB - k_2 B
\label{eqn:eqn2}
\end{equation}
where $B$ and $NB$ are a number of budding and non-budding cells respectively. It can be shown that the long-time behavior of the ratio of number of non-budding and budding cells reaches a stationary value ~\nameref{S1_Fig} and hence 
\begin{equation}
\frac{NB}{B}=r
\label{eqn:eqn3}
\end{equation}
where the ratio $r$ is related to the budding index by $r^{-1}=f_B/f_{NB}$ and depends only on the values of $k_1$ and $k_2$ and hence only the values of $\tau_1$ and $\tau_2$.  It is found the value of $r$ can be different for different strains of budding yeast \cite{Zettel2003}. The long-time behavior $r$ can be written in terms of $k_1$ and $k_2$ (see the Supporting Information)
\begin{equation}
r  = \frac{(k_2-k_1)+\sqrt{(k_1+k_2)^2+4k_1k_2}}{2k_1}
\label{eqn:eqn7}
\end{equation}
Hence the fraction of $\emph{f}_{NB}$, and $\emph{f}_B$ of the cells at non-budding state and the budding state at normal pressure, respectively can be written as
\begin{equation}
f_{NB}=\frac{r}{1+r}
\label{eqn:eqn8}
\end{equation}
 and 
\begin{equation}
f_{B}=\frac{1}{1+r}
\label{eqn:eqn9}
\end{equation}
\begin{figure}[htb]
	\begin{center}
		\includegraphics[width=12cm]{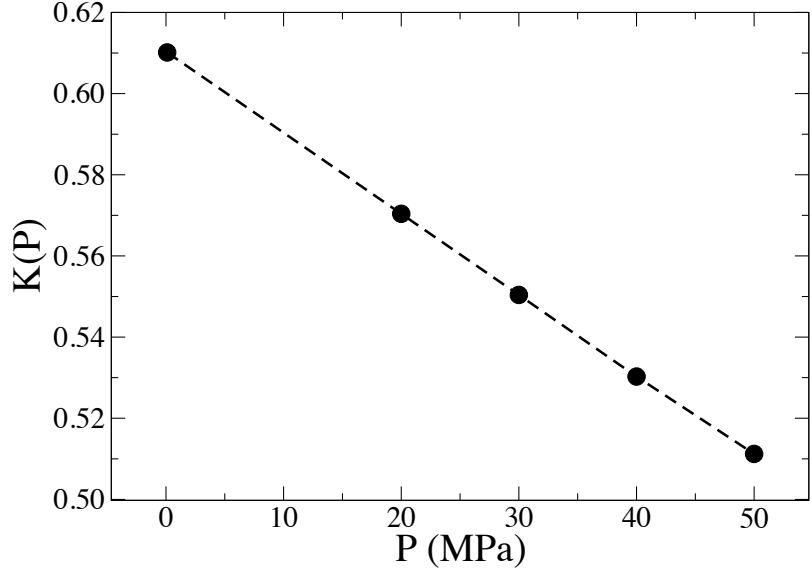}
		\caption{The ratio of $K(P)$ as a function of pressure. The ratio $K(P)$ decreases with increasing pressure leading to a decrease of budding index at high pressure.}
		\label{fig:Kp_K}
	\end{center}
\end{figure}

High hydrostatic pressure affects both the growth and budding index of yeast cells. A general continuum model that captures the effects of high pressure on budding yeast \textit{S.cerevisiae} is shown in Fig. \ref{fig:model} (b).  At high pressure, non-budding and budding cells have a small probability of going through cellular arrest and death, which inhibits them from proceeding through the normal cell cycle. We expect that the high pressure will affect both $k_1$ and $k_2$, as well as the likelihood of cellular arrest (or death) in the budding and non-budding states. The above argument leads to a modification of the growth kinetics of the cell division and can be written as:

\begin{equation}
\frac{dNB}{dt} = -(k_1+k_3)(NB) + 2k_2 B
\label{eqn:eqn10}
\end{equation}
\begin{equation}
\frac{dB}{dt} = k_1(NB) - (k_2+k_4)B
\label{eqn:eqn11}
\end{equation}
\begin{equation}
\frac{dNB_{CA}}{dt} = k_3NB
\label{eqn:eqn12}
\end{equation}

\begin{equation}
\frac{dB_{CA}}{dt} = k_4B
\label{eqn:eqn13}
\end{equation}
where $B$, $NB$, $B_{CA}$, $NB_{CA}$ are the number of budding, non-budding, cell cycle arrested budding, and cell cycle arrested non-budding cells respectively, and $k_3$ and $k_4$ are pressure-dependent rates of cellular arrest or death arising in non-budding and budding states, respectively. The ratio $r$, depends only on $k_1$ and $k_2$, is given by:

\begin{equation}
r  = \frac{(k_2(P)-k_1(P))+\sqrt{(k_1(P)+k_2(P))^2+4k_1(P)k_2(P)}}{2k_1(P)}
\end{equation}

In Fig. \ref{fig:Kp_K}, we show the behavior of the ratio of $K(P)=\frac{k_1}{k_2} (P)$ as a function of pressure, $P$. Our results show that $K(P)$ decreases upon increasing pressure, resulting in the budding index decrease with increasing pressure discussed above. Determination of parameters $k_3$, and $k_4$ will require further experiments quantifying the fraction of cell deaths at different pressures that we will explore in the future.

\section*{Discussion}

 Adaptive evolution of organisms to extremes of physicochemical conditions poses an attractive and exciting challenge to understand the evolutionary trajectories of adaptation. Before one can explore adaptive evolution of mesophilic organisms to extreme conditions, one must first investigate the effects of these conditions on the organism in order to determine the best parameter space for evolution experiments. We have performed a quantitative investigation of the effects of high hydrostatic pressure on the budding index, cell morphology, and cell death of a model eukaryote, \emph{Saccharomyces cerevisiae}.  We find that while cells are able to grow and reproduce up to $50$~MPa, a small fraction of cells either undergo cell death or cellular arrest in both budding and non-budding stages of cell cycle.  Furthermore budding index of the cells decreases with increasing pressure. Our phenomenological model that captures the cellular states -- namely budding and non-budding -- suggests that the ratio of rates determining the cell state changes from non-budding to budding ($k_1$) and from budding to non-budding ($k_2$) decreases with increasing pressure. Moreover, we find that the high pressure affects morphological determinants of the cells such as ellipticity and cell size. Our results show that cells become smaller and less elliptical upon increasing pressure. These changes in morphology may arise due to high pressure induced changes in cell wall elasticity.
 
 In addition, we find that high pressure induces both apoptotic and necrotic cell deaths albeit with a much smaller probability. Yeast cells can undergo apoptosis showing some specific markers, such as DNA cleavage and apoptosis-typical chromatin condensation (margination),  externalization of phosphatidylserine to the outer leaflet of the plasma membrane \cite{Madeo1997}. On the other hand, necrotic cells (accidental cell death) results from devastating cellular injuries such as chemical or physical disruption of the plasma membrane \cite{Eisenberg2010}, dysfunction of cell organelles, such as vacuole dysfunction, peroxisomal dysfunction. It has been shown that there is evidence that vacuole acidification at high pressure \cite{Abe1995} causes a decrease of pH inside the cells leading to cell death. However, we also find evidence of early and late apoptosis of the cells exposed to very high pressure. This could be caused by one of the many factors that cause DNA damage such chromatin condensation. Factors determining apoptotic and necrotic cell death at high pressure will be studied in details in our future experiments. The region of the parameter space relevant for evolution experiments presumably lies at the tipping point between cell growth and cell death. Our results suggest that for the  temperature $37^{\circ}$~C,  a pressure range of $50-70$~MPa  is a viable choice for  the adaptive evolution of \textit{S. Cerevisiae} to high pressures.

\section*{Conclusion}
We have performed quantitative investigations of the effects of pressure on \emph{Saccharomyces cerevisiae}. More specifically, we have quantified the effect of high hydrostatic pressure on the budding index, cell morphology, and cell viability of this model eukaryote. We find that high hydrostatic pressure decreases the budding index both via cell cycle arrest and cell death. While the majority of the cells are still able to undergo budding at high pressures up to $50$~MPa; there is a finite probability of cellular arrest and cell death as the pressure increases. Moreover, we find that the likelihood of cell death increases upon increasing pressure.  We find evidence of apoptotic and necrotic cell deaths upon increasing pressure with a higher proportion of necrotic death. Application of high hydrostatic pressure results in cell morphology changes as well. Specifically, we find that the cells become progressively smaller and less elliptical upon increasing pressure. Finally, we develop a phenomenological continuum model to account for the effects of pressure on the budding index. Our results besides providing invaluable quantitative insights into the effect of pressure on the budding cycle and cell death of \emph{S. cerevisiae}, paves the way for determining the region of temperature and pressure relevant for adaptive evolution of this model eukaryote, a theme that our lab has been exploring.

\section*{Supporting Information}

\paragraph*{S1}
\label{S1_Fig}
{\bf Long-time limit of the ratio of budding and non-budding cells in a population}.

Here we show that the ratio of non-budding to budding cells in a population reaches a constant in the long-time limit.  A schematic of the process is shown in figure ~\ref{fig:model}. We assume that the rate constant of initial growth of the unbounded cell is $k_1$ which is related to the time-scale of the initial growth $\tau_1$. Once a bud is formed, two new cells are formed from the budding cells  over a time-scale $\tau_2$ or with a rate constant $k_2$.

Following the above argument, we can write the growth kinetics of budding yeast as
\begin{equation}
\frac{dNB}{dt} = -k_1 NB + 2k_2 B
\end{equation}
\begin{equation}
\frac{dB}{dt} = k_1 NB - k_2 B
\end{equation}
In the following, we will show that the ratio of the budding to non-budding cells at long times reaches a stationary value whereas this ratio depends on $k_1$ and $k_2$ and is independent of initial values of $NB$ and $B$.  

Dividing Eq.~1 by $1/B$ and Eq.~2 by $NB/B^2$ and subtracting the resulting equations, we get
\begin{equation}
\frac{1}{y}\frac{dNB}{dt}-\frac{NB}{B^2}\frac{dB}{dt} = -k_1(\frac{NB}{B}) + 2k_2 -k_1(\frac{NB}{B})^2+k_2(\frac{NB}{B})
\end{equation}
Substituting $r=\frac{NB}{B}$, which is the ratio of non-budding to budding cells, we get
\begin{equation}
\frac{dr}{dt} = (k_2-k_1)r - k_1r^2+2k_2
\end{equation}
Solution of the above equation can be written as 
\begin{equation}
r(t) = \frac{A- B\psi e^{-\frac{k_1}{A-B}t}}{1-\psi e^{-\frac{k_1}{A-B}t}}
\end{equation}
where $A=\frac{(k_2-k_1)+\sqrt{(k_1+k_2)^2+4k_1k_2}}{2k_1}$, $B=-\frac{4k_2}{(k_1-k_2)+\sqrt{(k_1+k_2)^2+4k_1k_2}}$, and $\psi=\frac{r(t=0)-A}{r(t=0)-B}$ Since, $A-B>0$, the long-time limit of $r(t)$ converges and hence
\begin{equation}
\lim_{t \rightarrow \infty} r(t) = \lim_{t \rightarrow \infty} \frac{A- B\psi e^{-\frac{k_1}{A-B}t}}{1-\psi e^{-\frac{k_1}{A-B}t}} =  A=\frac{(k_2-k_1)+\sqrt{(k_1+k_2)^2+4k_1k_2}}{2k_1}
\end{equation}

\section*{Acknowledgments}

We thank Sudip Nepal, Kionna Henderson, and Barrett Johnson for fruitful discussions and Arkansas Bio Institute for support.


\begin{thebibliography}{10}


\bibitem{Yayanos1981} Yayanos AA, Dietz AS, Boxtel RV. Obligately barophilic bacterium from the Mariana trench. Proc. Natl. Acad. Sci. USA. 1981;78: 5212-5215.

\bibitem{Kato1998} Kato C, Li L, Nogi Y, Nakamura Y, Tamaoka J, Horikoshi K. Extremely barophilic bacteria isolated from the Mariana Trench, Challenger Deep, at a depth of 11,000 meters. Appl. Environ. Microbiol. 1998;64: 1510-1513.

\bibitem{Brock1969} Brock TD, Freeze H. Thermus aquaticus gen. n. and sp. n., a nonsporulating extreme thermophile. J. Bacteriol. 1969;98: 289-297.

\bibitem{Bakermans2003} Bakermans C, Tsapin AI, Souze-Egipsy V, Gilichinsky DA, Nealson KH. Reproduction and metabolism at $-10^o$C of bacteria isolated from Silberian permafrost. 2003;5: 321-326.

\bibitem{Schleper1995} Schleper C, Piihler G, Kuhlmorgen B, Zillig W. Life at extremely low pH. Nature. 1995;375: 741-742.

\bibitem{Horikoshi1982} Horikoshi K, Akiba T. Alkalophilic Microorganisms: A New Microvial World. Springer: Heidelberg, Germany. 1982.  

\bibitem{Hawley1971} Hawley SA. Reversible pressure-temperature denaturation of chymotrypsinogen. Biochemistry. 1971;10: 2436-2442. 

\bibitem{Mozhaev1996} Mozhaev VV, Heremans K, Frank J, Masson P, Balny C. High pressure effects on protein structure and function. Proteins: Structure, Function and Bioformatics. 1996;24: 81-91.

\bibitem{Privalov1990} Privalov PL. Cold denaturation of proteins. Crit. Rev. Biochem Mol. Biol. 1990;25: 281-305.


\bibitem{Gross1994} Gross M, Jaenicke R. Proteins under pressure. The influence of high hydrostatic pressure on structure, function and assembly of proteins and protein complexes. Eur. J. Biochem. 1994;221: 617-630.

\bibitem{Hummer1998} Hummer G, Garde S, Garcia AE, Paulaitis ME, Pratt LR. The pressure dependence of hydrophobic interactions is consistent with the observed pressure denaturation of proteins. PNAS. 1998;95: 1552-1555.

\bibitem{Merrin2011} Merrin J, Kumar P, Libchaber A. Effects of pressure and temperature on the binding of RecA protein to single-stranded DNA. PNAS. 2011;108: 19913-19918.

\bibitem{Kumar2013} Kumar P, Lichaber A. Pressure and temperature dependence of growth and morphology of \textit{Escherichia coli}: Experiments and Stochastic Model. Biophysic Journal. 2013;105: 783-793.

\bibitem{Abe2004} Abe F. Piezophysiology of yeast: occurrence and significance. Cellular and Molecular Biology 2004;50: 437-445.

\bibitem{Sonoike1992} Sonoike K, Setoyama T, Kuma Y, Shinno T, Fukumoto K, Ishihara M. Effects of pressure and temperature on death rate of \emph{Lactobacillus casei} and \emph{Escherichia coli}. In: Blany C, Hayashi R, Heremans K, Masson P, editors. High Pressure and Biotechnology. Colloque INSERM. 1992;224: 297-301.  

\bibitem{Seki1998} Seki K, Toyoshima M. Preserving tardigrades under pressure. Nature. 1998;395: 853-854.

\bibitem{Ishii2004} Ishii A, Sato T, Kato C. Effects of high hydrostatic pressure on bacterial cytoskeleton FtsZ polymers in vivo and in vitro. Microbiology. 2004;150: 1965-1972.

\bibitem{Kumar2016} Kumar P, Libchaber A. Cell fate and reversibility of \textit{Escherichia coli} with pressure. (under review).

\bibitem{Huh2011} Huh D, Paulsson J. Random partitioning of molecules at cell divisions. PNAS. 2011;108: 15004-15009.

\bibitem{Nishiyama2012} Nishiyama M, Sowa Y. Microscopic analysis of bacterial motility at high pressure. Biophysical Journal. 2012;102: 1872-1880.

\bibitem{Hartwell1974} Hartwell L. \emph{Saccharomyces cerevisiae} Cell Cycle. Bacteriological Reviews. 1974;38: 167-1980.

\bibitem{Johnston1977} Johnston GC, Pringle JR, Hartwell LH. Coordination of growth with cell division in the yeast \emph{Saccharomyces cerevisiae}. Experimental Cell Research 1977;{105}: 79-98.

\bibitem{Hartwel1974} Hartwell LH, Culotti J, Pringle JR, Reid BJ. Genetic control of the cell division in yeast. Science 1974:{183}: 46-51.

\bibitem{Fernandes2005} Fernandes PM. How does yeast response to pressure? Brazilian Journal of Medical and Biological Research 2005;38: 1239-1245.

\bibitem{Palou1998} Palou E, Lopez-Malo A, Barbosa-Canovas GV, Welti=Chanes J, Davidson PM, Swanson G. High hydrostatic pressure come up time and yeast viability. Journal of Food Protection. 1998;{61}: 1657-1660.

\bibitem{Shibata2014} Shibata M, Torigoe M,  Matsumoto Y, Yamamoto, M, Takizawa N, Hada Y, et al. Tolerance of budding yeast {\it Saccharomyces cerevisiae} to ultra high pressure. Journals of Physics: Conference Series. 2014;500. 

\bibitem{Lammi2004} Lammi M, Elo M, Sironen R, Karjalainen H, Kaarniranta K, Helminen H. Hydrostatic pressure-induced changes in cellular protein synthesis. Biorheology. 2004;41: 309-313.

\bibitem{Rosin1977} Rosin MP, Zimmerman AM. The induction of cytoplasmic petite mutants of \emph{Saccharomyces cerevisiae} by hydrostatic pressure. J. Cell Sci. 1977;26: 373-385.

\bibitem{Hamada1992} Hamada K, Nakatomi Y, Shimada S. Direct induction of tetraploids or homozygous diploids in the industrial yeast \emph{Saccharomyces cerevisiae} by hydrostatic pressure. Cure Genet. 1992;22: 371-376.

\bibitem{Bravim2014} Bravim F, Silva L, Souza D, Lippman S, Broach J, Fernandes A, Fernandes P. High hydrostatic pressure activates transcription factors involved in {\it Saccharomyces cerevisiae} stress tolerance. Curr Pharm Biotechnol. 2014;13: 2712-2720 .

\bibitem{Fernandes2004} Fernandes P, Domitrovic T, Kao C, Kurtenbach E. Genomic expression pattern in {\it Saccharomyces cerevisiae} cells in response to high hydrostatic pressure.  FEBS Letters. 2004;556: 153-160.

\bibitem{Kobori1995} Kobori H, Sato M, Tameike A, Shimada S, Osumi M. Ultrastructure effects of pressure stress to the nucleus in \emph{Saccharomyces cerevisiae}: a study by immunoelectron microscopy using frozen thin sections. FEMS Microbiol. Lett. 1995;132: 253-258.

\bibitem{Iwahashi1991} Iwahashi H, Kaul SC, Obuchi K, Komatsu Y. Induction of barotolerance by heat shock treatment in yeast. FEMS Microbiol. Lett. 1991;80: 325-328.

\bibitem{Kim2007} Kim HS, Fay JC. Genetic variation in the cysteine biosynthesis pathway causes sensitivity to pharmacological compounds. Proc. Natl. Acad. Sci. USA. 2007;104: 19387-19391. 

\bibitem{Saghbini2001} Saghbini M, Hoekstra D, Gautsch J. Media formulations for various two-hybrid systems. Methods Mol. Biol. 2001;177: 15-39.

\bibitem{Gal2003} Gal O. Fit Ellipse. MathWorks: in MatLab. Available from: https://www.mathworks.com/matlabcentral/fileexchange/3215-fit-ellipse. 

\bibitem{Suzuki2004} Suzuki M, Asada Y, Watanabe D, Ohya Y. Cell shape and growth of budding yeast cells in restrictive microenvironments. {Yeast}. 2004:{21}: 983-989.

\bibitem{Becze1995} Becze GE. A Microbiological Process Report Yeast: I. Morphology {Applied and Environmental Microbiology}. 1995.

\bibitem{Knop2011} Knop M. Yeast cell morphology and sexual reproduction - a short overview and some considerations. {Comptes Rendus Biologies}. 2011;{334}: 599-606.

\bibitem{Madeo1997} Madeo F, Frohlich E, Frohlich K. A yeast mutant showing diagnostic markers of early and late apoptosis. {The Journal of Cell Biology.} 1997;{139}: 729-734.

\bibitem{Gutierrez2010} Carmona-Gutierrez D, Eisenberg T, Buttner S, Meisinger C, Kroemer G, Madeo F. Apoptosis in yeast: triggers, pathways, subroutines. Cell Death and Differentiation. 2010;{17}: 763-773.

\bibitem{Eisenberg2010} Eisenberg T, Carmona-Gutierrez D, Buttner S, Tavernarakis N, Madeo,F. Necrosis in yeast. {Apoptosis.} 2010;{15}: 257-268.

\bibitem{Buttner2006} Buttner S, Eisenberg E, Herker E, Carmona-Gutierrez D, Kroemer G, Madeo F. Why yeast cells can undergo apoptosis: death in time of peace, love and war. {The Journal of Cell Biology.} 2006;{175}: 521-525.


\bibitem{Gutierrez2009} Carmona-Gutierrez D, Madeo F. Tracing the roots of death: apoptosis in \emph{Saccharomyces cerevisiae}. In: Zheng D, Yin XM, editors. Essentials of Apoptosis: A guide for basic and clinical research. {Humana Press}; 2009. pp. 325-354.

\bibitem{Zong2006} Zong W, Thompson CB. Necrotic death as cell fate. {Genes \& Dev.} 2006;{20}: 1-15.

\bibitem{Zettel2003} Zettel MF, Garza LR, Cass AM, Myhre RA, Haizlip LA, Osadebe SN, et al. The budding index of \textit{Saccharomyces cerevisiae} deletion strains identifies genes important for cell cycle progression. FEMS Micro. Lett. 2003;223: 253-258.


\bibitem{Abe1995} Abe F, Horikoshi K. Hydrostatic pressure promotes the acidification of vacuoles in \emph{Saccharomyces cerevisiae}. FEMS Microbiol. Lett. 1995;{130}: 307-312.


\end{thebibliography}
\end{document}